\documentclass[12pt]{article}

\usepackage{slashed}
\usepackage{latexsym}
\usepackage{amsfonts}
\usepackage{cite}
\usepackage{amsmath}

\def\nn{\nonumber}

\csname @addtoreset\endcsname{equation}{section}
\newcommand{\ft}[2]{{\textstyle\frac{#1}{#2}}}
\def\rmi{{\,\rm i\,}}
\newsavebox{\uuunit}
\sbox{\uuunit}
    {\setlength{\unitlength}{0.825em}
     \begin{picture}(0.6,0.7)
        \thinlines
        \put(0,0){\line(1,0){0.5}}
        \put(0.15,0){\line(0,1){0.7}}
        \put(0.35,0){\line(0,1){0.8}}
       \multiput(0.3,0.8)(-0.04,-0.02){12}{\rule{0.5pt}{0.5pt}}
     \end {picture}}

\def\be{\begin{equation}}
\def\ee{\end{equation}}
\def\ba{\begin{array}}
\def\ea{\end{array}}
\def\bea{\begin{eqnarray}}
\def\eea{\end{eqnarray}}
\def\bd{\begin{displaymath}}
\def\ed{\end{displaymath}}

\def\nn{\nonumber}
\def\qq{\quad\quad}

\def\g{\gamma}

\def\d{\delta}

\def\e{\epsilon}
\def\ve{\varepsilon}
\def\f{\phi}

\def\p{\psi}

\def\l{\lambda}

\def\m{\mu}
\def\n{\nu}
\def\r{\rho}
\def\s{\sigma}

\def\o{\omega}

\newcommand{\SU}{\mathop{\rm SU}}

\def\del{\partial}
\newcommand{\w}[1]{\\[0.#1cm]}

\begin{document}


\begin{center}

\hfill UG-11-77 \\
\hfill MIFPA-11-31

\vskip 1.5cm

{\Large \bf Off-shell D=5, N=2 Riemann Squared Supergravity
 }
\\

\vskip 2cm

{\bf Eric A.~Bergshoeff\,$^1$, Jan Rosseel\,$^1$  and  Ergin Sezgin\,$^2$}\\

\vskip 25pt

{\em $^1$ \hskip -.1truecm Centre for Theoretical Physics, University of Groningen,
\\
Nijenborgh 4, 9747 AG Groningen, The Netherlands \vskip 5pt }

{email: {\tt E.A.Bergshoeff@rug.nl, j.rosseel@rug.nl}} \\

\vskip 15pt

{\em $^3$ \hskip -.1truecm George and Cynthia Woods Mitchell Institute
for Fundamental Physics and Astronomy, Texas A\& M University,
College Station, TX 77843, USA}

{email: {\tt sezgin@tamu.edu}}
\\

\end{center}

\vskip 0.5cm

\begin{center} {\bf ABSTRACT}\\[3ex]
\end{center}

We construct a new off-shell invariant in $\mathcal{N}=2, D=5$
supergravity whose leading term is the square of the Riemann tensor.
It contains a gravitational Chern-Simons term involving the vector
field that belongs to the supergravity multiplet. The action is
obtained by mapping the transformation rules of a spin connection
with bosonic torsion and a set of curvatures to the fields of the
Yang-Mills multiplet with gauge group $\text{SO}(4,1)$.  We also
employ the circle reduction of an action that describes locally
supersymmetric Yang-Mills theory in six dimensions.


\thispagestyle{empty}

\newpage


\section{Introduction}

It is well-known that in string theory the Einstein action of
general relativity gets modified by an infinite series of terms of
higher order in the Riemann tensor. To obtain the supersymmetric
extension of such an infinite series is an open problem. In lower
dimensions there exist auxiliary field formulations of Poincar\'e
supergravity which makes it possible to construct supersymmetric
actions that contain only the first order correction to the Einstein
action. This has been done for $\mathcal{N}=2$ supersymmetry in
$D=6$ dimensions a long time ago
\cite{Bergshoeff:1986vy,Bergshoeff:1986wc,Bergshoeff:1987rb}. The
resulting action contains an Einstein plus a Riemann tensor squared
term. The construction of
\cite{Bergshoeff:1986vy,Bergshoeff:1986wc,Bergshoeff:1987rb} was
based on the observation that the Weyl multiplet underlying
Poincar\'e supergravity has formally the same supersymmetry
transformation rules as a Yang-Mills multiplet, for Yang-Mills group
$SO(5,1)$. Actually, there exist two different Weyl multiplets
\cite{Bergshoeff:1985mz} and this analogy only works for the
so-called Dilaton Weyl multiplet which contains, amongst others, a
dilaton-like scalar field that can be used as the compensating field
for dilatations and an antisymmetric tensor gauge field. In this
analogy the Yang-Mills vector $A_\mu^I$ transforms formally the same
as a certain torsionful spin connection $\omega_{\mu}{}^{ab}_+$
where the torsion is proportional to the three-form field-strength
tensor of the antisymmetric tensor gauge field. The supersymmetry
rules only coincide after fixing the conformal symmetries using the
scalar field to fix the dilatations.

Another example of a supersymmetric higher order invariant has been
constructed in $D=5$ dimensions \cite{Hanaki:2006pj}. The leading
term of this invariant is a Weyl curvature squared term that is
multiplied by a compensating scalar, to make it invariant under
dilatations. This compensating scalar belongs to a separate gauge
multiplet. An interesting feature of this invariant is that it
contains a mixed gauge-gravitational Chern-Simons term
\begin{equation}\label{mixed}
A\wedge \text{tr}( R\wedge R)\ ,
\end{equation}
where $A$ belongs to the gauge multiplet and $R$ to the supergravity
multiplet. This mixed Chern-Simons term plays an important role in
discussing higher-order corrections to black hole entropy, see
e.g.~\cite{Cremonini:2008tw,deWit:2009de}, and higher-order effects
in the AdS/CFT correspondence, see
e.g.~\cite{Liu:2010gz,Cremonini:2009sy}.

In this note we show that, by  applying the techniques of \cite{Bergshoeff:1986vy,Bergshoeff:1986wc,Bergshoeff:1987rb}
to $\mathcal{N}=2$ supersymmetry in $D=5$ dimensions we can construct a higher order invariant that differs from the one presented in \cite{Hanaki:2006pj}. The leading term of this new invariant is the
Riemann tensor squared. Unlike the invariant of \cite{Hanaki:2006pj} this one is purely gravitational
in the sense that the compensating scalar for dilatations, that multiplies the Riemann tensor squared term, belongs to the Weyl multiplet. In analogy to \eqref{mixed} the new invariant contains a purely gravitational Chern-Simons term
\begin{equation}\label{pure}
C\wedge \text{tr} (R\wedge R)\,,
\end{equation}
where both $C$ and  $R$ belong to the supergravity multiplet. It is
natural to expect that this term will also be relevant in exploring
the effects of higher-derivative corrections in black hole entropy
and the AdS/CFT correspondence.

This paper is organized as follows. In section 2 we will record the
relevant elements of the so-called Dilaton Weyl multiplet and
Yang-Mills multiplet with ${\cal N}=2$ supersymmetry in $5D$. In
section 3, we shall go over to a convenient basis for the fields
which is equivalent to fixing the superconformal symmetries, in the
sense that the new fields are invariant under dilatations and
$S$-supersymmetry. Section 4 contains the key observation of this
paper which states that the transformation rules of a spin
connection with bosonic torsion and a set of curvatures in the
Dilaton Weyl multiplet formally transform in the same manner as the
fields of a  Yang-Mills multiplet with gauge group $\text{SO}(4,1)$.
The explicit correspondence can be found in eq.~\eqref{rules}. After
we dimensionally reduce a locally supersymmetric Yang-Mills action
from $6D$ to $5D$ in section 5, we make use of this observation to
write down the supersymmetrization of the Riemann squared term in
section 7.  Further directions and comments are presented in the
concluding section.


\section{Conformal Multiplets}


In this section, we will briefly recall some elements of the
$\mathcal{N}=2$ superconformal tensor calculus in five dimensions,
that will be useful in the construction of the new higher-derivative
supergravity invariant. More specifically, we will review the
relevant Weyl multiplet containing the various gauge fields of the
superconformal symmetries and the Yang-Mills multiplet.   Most of
the results presented in this and the next section can be found in
\cite{Bergshoeff:2001hc,Bergshoeff:2002qk}.


\subsection{The Dilaton Weyl Multiplet}


There exist two Weyl multiplets in five dimensions, known as the
Standard Weyl multiplet and the Dilaton Weyl multiplet. They were
constructed in \cite{Bergshoeff:2001hc} and contain the same number
of gauge fields but differ in their matter field content. The
multiplet that is relevant for this paper is the Dilaton Weyl
multiplet. It consists of the vielbein $e_\m{}^a$, the gravitino
$\psi_\m^i$, the dilation gauge field $b_\mu$ and the $\SU(2)$ gauge
field $V_\mu^{ij} = V_\mu^{(ij)}$. These gauge fields are
supplemented with matter fields to form a multiplet consisting of
$32$ bosonic and $32$ fermionic off-shell degrees of freedom. For
the Dilaton Weyl multiplet, these matter fields are given by a
vector $C_\m$, an anti-symmetric tensor $B_{\m \n}$, a dilaton field
$\s$ and a fermion field $\p^i$. The $Q$- and $S$-supersymmetry
transformations (with parameters $\e^i$, $\eta^i$ respectively) are given
by
\allowdisplaybreaks
\begin{eqnarray}\label{weyl}
\d e_\m{}^a  &=&  \ft 12\bar\e \g^a \psi_\m  \nn\, ,
\\[.2truecm]
\d \psi_\m^i   &=& D_\m (\widehat{\omega}) \e^i + \rmi \g\cdot T \g_\m
\e^i - \rmi \g_\m \eta^i  \nn\, ,
\\[.2truecm]
\d V_\m{}^{ij}  &=&  -\ft32\rmi \bar\e^{(i} \phi_\m^{j)}
+4\bar\e^{(i}\g_\m \chi^{j)}+ \rmi \bar\e^{(i} \g\cdot T \psi_\m^{j)} + \ft32\rmi
\bar\eta^{(i}\psi_\m^{j)} \nn\, ,
\\[.2truecm]
\d C_\m &=& -\ft12\rmi \s \bar{\e} \p_\m + \ft12
\bar{\e} \g_\m \p 
\, , \nonumber
\\[.2truecm]
\d B_{\m\n} &=& \ft12 \s^2 \bar{\e} \g_{[\m} \p_{\n]}
+ \ft12 \rmi \s \bar{\e}\g_{\m\n} \p + C_{[\m} \d(\e) C_{\n]}
\, , \nonumber
\\[.2truecm]
\d \p^i  &=& - \ft14 \g \cdot \widehat{G} \e^i -\ft12\rmi
\widehat{\slashed{D}} \s \e^i + \s \g \cdot T \e^i - \ft14 \rmi \s^{-1} \e_j
\bar{\p}^i \p^j + \s\eta^i \, , \nonumber
\\[.2truecm]
\d \s &=& \ft12 \rmi \bar{\e} \p \, ,\nonumber
\\[.2truecm]
\d b_\m &=& \ft12 \rmi \bar\e\phi_\m -2 \bar\e\g_\mu \chi +
\ft12\rmi \bar\eta\psi_\mu \,. 
\label{weyl}
\end{eqnarray}
The `soft' algebra that the Dilaton Weyl multiplet realizes is given in \cite{Bergshoeff:2001hc}.
Several definitions, some of which will be needed later, are as follows. Firstly,
\begin{eqnarray}\label{cd}
 D_\m (\widehat{\omega}) \e^i & = & \partial_\m \e^i+\ft 12 b_\m\e^i +\ft 14 \widehat{\o}_\m^{ab}
\g_{ab}\e^i - V^{ij}_\m\epsilon _j\ ,
\nn\w2
 \widehat{D}_\m \sigma & = & \partial_\mu\sigma -b_\mu \s -\tfrac12 \rmi {\bar\psi}_\mu\psi\ ,
 \nn\w2
 \widehat{D}_\mu \psi^i &=&  (\partial_\mu - \tfrac32 b_\mu +\tfrac14 \widehat{\omega}_\mu{}^{ab} \g_{ab})\psi^i -V_\mu^{ij}\psi_j
 +\tfrac14 \g\cdot \widehat{G} \psi_\mu^i + \tfrac12 \rmi \widehat{\slashed{D}}\sigma \psi_\mu^i\nn
 \\[.2truecm]
 && -\sigma \g\cdot T\psi_\mu^i
 +\tfrac14 \rmi \sigma^{-1} \psi_{\mu j} {\bar\psi}^i\psi^j-\sigma \phi_\mu^i\ .
\end{eqnarray}
Moreover, in the transformation rules we have used the composite fields
\begin{eqnarray}
T_{ab} &=& \ft18 \s^{-2}\bigg(\s \widehat{G}_{ab} + \ft16
\ve_{abcde} \widehat{H}^{cde} + \ft14\rmi \bar{\p} \g_{ab} \p \bigg)
\, ,\label{TchiindilW}
\nn\w2\nn
\chi^i &=& \ft18\rmi \s^{-1} \widehat{\slashed{D}} \p^i +\ft1{16}\rmi \s^{-2}
\widehat{\slashed{D}} \s \p^i - \ft1{32} \s^{-2} \g \cdot \widehat{G} \p^i
\nn\w2
& & + \ft14 \s^{-1} \g \cdot T \p^i +
\ft1{32}\rmi \s^{-3}  \p_j\bar{\p}^i \p^j \ .
\label{valueD}
\eea
In fact, $T_{ab}, \chi$ and a scalar field $D$, which does not arise
here, constitute the matter fields of the so-called Standard Weyl
multiplet \cite{Bergshoeff:2001hc}, and the above expressions are
needed to pass from this to the Dilaton Weyl multiplet. The
expressions for the dependent spin connection
${\widehat\omega}_\mu{}^{ab}$ and the $S$-supersymmetry gauge field
$\phi_\mu^i$ are given by
\bea
\widehat{\omega}^{ab}_\mu
&=& 2 e^{\nu[a} \partial_{[\mu} e_{\nu]}^{~b]} - e^{\nu[a} e^{b]\s} e_{\mu c}
\partial_\nu e^{~c}_\s
 + 2 e_\mu^{~~[a} b^{b]} +\ft12 \bar{\psi}^{[a} \gamma^{b]} \psi_\mu + \ft14
\bar{\psi}^a \gamma_\mu \psi^b ,
\nn\w2
\f^i_\m &=& \ft13\rmi \g^a \widehat{R}^\prime _{\m a}{}^i(Q) -
\ft1{24}\rmi \g_\m \g^{ab} \widehat{R}^\prime _{ab}{}^i(Q)\ .
\label{deff}
\eea
We have the field strengths for $C_\mu, B_{\mu\nu}$ and $
V_\mu^{ij}$ defined as
\bea
\widehat{G}_{\m\n} &=& 2\partial _{[\mu }C_{\nu]} + \ft12\rmi \s
\bar{\p}_{[\m} \p_{\n]}
 -  \bar{\p}_{[\m} \g_{\n]} \p\ ,
\label{hatG}\w2
\widehat{H}_{\m\n\r} &=&3\partial _{[\mu }B_{\nu \rho ]} + \ft32 C_{[\m} G_{\n\r]} - \ft34
\s^2 \bar{\p}_{[\m} \g_\n \p_{\r]} - \ft32\rmi \s \bar{\p}_{[\m}
\g_{\n\r]} \p \ ,
\label{hatFH}\w2
{\widehat R}_{\mu\nu}{}^{ij}(V) &=& 2\partial_{[\mu}V_{\nu]}{}^{ij} -2V_{[\mu}{}^{k(i}V_{\nu]k}{}^{j)} -3i{\bar\phi}^{(i}_{[\mu}\psi_{\nu]}^{j)}-8{\bar\psi}_{[\mu}^{(i}\gamma_{\nu]}\chi^{j)}
-i{\bar\psi}_{[\mu}^{(i} \gamma\cdot T\psi_{\nu]}^{j)}\ ,
\label{RV}
\eea
and the gravitino curvature as
\bea \widehat{R}_{\m\n}{}^i(Q) &=& \widehat{R}^\prime
_{\mu\nu}{}^i(Q) - 2\rmi\g_{[\m} \f_{\n]}^i\ , \label{curv1}\w2
{\widehat R}^\prime _{\mu\nu}{}^i(Q)  &=&  2\del_{[\m}\p^i_{\n ]}
+\ft12 \widehat{\o}_{[\m}{}^{ab} \g_{ab}\p^i_{\n ]} +b_{[\m}\p^i_{\n
]} -2V_{[\m}{}^{ij}\p_{\n ]\,j}+2\rmi \g\cdot T \g_{[\mu}
\psi_{\nu]}^i\ . \label{curv2} \eea
Note that ${\widehat H}_{\mu\nu\rho}$ is invariant under the bosonic gauge transformations
\be \delta C_\mu= \partial\Lambda\ ,\qquad \delta B_{\mu\nu} =
2\partial_{[\mu} \Lambda_{\nu]} -\ft12 \Lambda G_{\mu\nu}\ ,\ee
where $G_{\mu\nu} = 2\partial_{[\mu}C_{\nu]}$.  Finally, for future
reference we give the transformation rules
\begin{eqnarray}
\d \widehat{\o}_\m{}^{ab} &=& \ft12\rmi \bar\e \g^{ab}\phi_\m  - \ft12\rmi \bar
\eta \gamma ^{ab}\psi _\mu  - \rmi \bar\e \g^{[a} \g\cdot T \g^{b]} \psi_\m \nn\\[.2truecm]
 && -\ft 12 \bar\e\g^{[a} \widehat R_\m{}^{b]}(Q)
 - \ft 14 \bar\e\g_\m \widehat R^{ab}(Q)  - 4 e_\m{}^{[a} \bar\e \g^{b]} \chi\ ,
\label{deltaO}\w2
\d \widehat{H}_{abc}
&=& -\ft34 \s^2 \bar{\e} \g_{[a} \widehat{R}_{bc]}(Q) + \ft32\rmi \bar{\e}
\g_{[ab} \widehat{D}_{c]} \p + \ft32 \rmi  \bar{\e} \g_{[ab]}\p \widehat{D}_{c]} \s
\nn \\[.2truecm]
& & - \ft32 \s \bar{\e} \g_{[a} \g \cdot T \g_{bc]} \p -\ft32
\bar{\e} \g_{[a} \widehat{G}_{bc]} \p  - \ft32 \s \bar{\eta} \g_{abc} \p \ ,
\label{deltaH}\w2
\d \widehat{G}_{ab}
&=& - \ft12\rmi \s \bar{\e} \widehat{R}_{ab}(Q) - \bar{\e} \g_{[a} \widehat{D}_{b]} \p
 + \rmi \bar{\e} \g_{[a} \g \cdot T \g_{b]} \p  +\rmi \bar{\eta} \g_{ab} \p \ .
\label{deltaF}
\end{eqnarray}
This concludes our short review of the Dilaton Weyl multiplet.


\subsection{The Yang-Mills Multiplet}


The off-shell non-abelian $D=5$, $\mathcal{N}=2$ vector multiplet
consists of $8 n + 8 n$ bosonic and fermionic degrees of freedom
(where $n$ denotes the dimension of the gauge group). Denoting the
Yang-Mills index by $I$ ($I=1,\cdots,n$), the bosonic field content
consists of vector fields $A_\m^I$, scalar fields $\rho^I$ and
auxiliary fields $Y^{ij\, I} = Y^{(ij)\, I}$, that are
$\SU(2)$-triplets. The fermion fields are given by $\SU(2)$-doublets
$\l^{i I}$.

The $Q$- and $S$-transformations of the vector multiplet, in the
background of the Dilaton Weyl multiplet, are given by \cite{Bergshoeff:2002qk}
\begin{eqnarray}
\d A_\m^I &=& -\ft12\rmi \rho^I \bar{\e} \p_\m + \ft12 \bar{\e}
\g_\m \lambda^I \ ,
\nonumber \w2
\d Y^{ij \, I} &=& -\ft12
\bar{\e}^{(i} \widehat{\slashed{D}} \lambda^{j) I} + \ft12 \rmi \bar{\e}^{(i}
\g \cdot T \lambda^{j) I} - 4 \rmi \rho^I \bar{\e}^{(i} \chi^{j)} +
\ft12 \rmi \bar{\eta}^{(i} \lambda^{j) I} - \ft12 \rmi g
\bar{\e}^{(i} f_{JK}{}^I \rho^J \lambda^{j)K} \ ,
\nonumber \w2
\d\lambda^{i I} &=& - \ft14 \g \cdot \widehat{F}^I \e^i -\ft12\rmi
\widehat{\slashed{D}}\rho^I \e^i + \rho^I \g \cdot T \e^i - Y^{ij\, I} \e_j +
\rho^I \eta^i \ ,
\nonumber \w2
\d \rho^I &=& \ft12 \rmi \bar{\e}
\lambda^I \ . \label{vectconform}
\end{eqnarray}
We have used here the superconformally covariant derivatives
\begin{eqnarray}
{\widehat D}_\mu\, \rho^I &=&
(\partial_\mu - b_\mu) \rho^I+ g f_{JK}{}^I A_\mu^J \rho^K
- \frac12\, \rmi\bar{\psi}_\mu \lambda^I \ ,
\label{deriv1}\w2
{\widehat D}_\mu \lambda^{iI} &=&
(\partial_\mu -\ft32 b_\mu +\ft14\,  \widehat{\o}_\mu
{}^{ab}\g_{ab} ) \l^{iI} - V_\mu^{ij} \l_j^I + g f_{JK}{}^I A_\mu^J
\lambda^{iK}
\nn\\
&& +\frac 14 \g
\cdot \widehat{F}^I \p_\mu^i + \ft12\rmi \widehat{\slashed{D}}\rho^I \p_\mu^i
 + Y^{ijI} \p_{\mu\, j}- \rho^I \g \cdot T \p_\mu^i - \rho^I
\phi_\mu^i \label{deriv2} \ ,
\end{eqnarray}
and the supercovariant Yang-Mills curvature
\begin{equation}
\widehat{F}_{\m\n}^I = 2 \partial_{[\mu } A_{\nu ]}^I + g f_{JK}{}^I
A_\mu^J A_\nu^K - \bar{\p}_{[\m} \g_{\n]} \lambda^I + \frac 12 \rmi
\rho^I \bar{\p}_{[\m} \p_{\n]}\ .
\label{hatF}
\end{equation}
%


\section{Change of Basis}


In what follows, it turns out to be convenient to change to a basis,
denoted by tilded fields, in which all fields are  dilatation and
$S$-supersymmetry invariant. In terms of the original fields the
tilded fields are given by
\begin{eqnarray}
{\tilde e}_\mu^a  &=& \sigma e_\mu^a\ ,
\nn\w2
{\tilde\p}_\mu^i &=& \sigma^{1/2} \p_\mu^i +\rmi \sigma^{-1/2}\g_\mu\psi^i\ ,
\nn\w2
{\tilde V}_\mu^{ij}  &=& V_\mu^{ij}-\ft32 \rmi \sigma^{-1}{\bar\p}_\mu^{(i}\psi^{j)}
+\ft34 \sigma^{-2}{\bar\p}^{(i}\g_\mu \psi^{j)}\ ,
\nn\w2
{\tilde B}_{\mu\nu} &=& B_{\mu\nu}\ ,
\nn\w2
{\tilde C}_\mu  &=& C_\mu \ ,
\nn\w2
{\tilde\epsilon} &=& \sigma^{1/2} \epsilon\ .
\label{redefs}
\end{eqnarray}
Dropping the tildes for convenience in notation, we find that the supersymmetry transformation
rules in the  new basis are given by
\begin{eqnarray}\label{fixed}
\d e_\m{}^a  & = &   \ft 12\bar\e \g^a \psi_\m \ , \nn
\\[.2truecm]
\d \psi_\m^i  & = & D_\m (\widehat{\omega}_-) \e^i -\ft12 \rmi
{\widehat G}_{\mu\nu} \g^\nu\e^i \ , \nn
\\[.2truecm]
\d V_\m{}^{ij} & = & \ft12 \bar\e^{(i} \g^\nu \psi_{\mu\nu}^{j)}
-\ft16 \bar\e^{(i} \g\cdot {\widehat H}\psi_\mu^{j)} -\ft14 \rmi
\bar\e^{(i} \g\cdot {\widehat G}\psi_\mu^{j)}\ , \nn
\\[.2truecm]
\d C_\m & = & -\ft12\rmi \bar{\e} \p_\m \ , \nn
\\[.2truecm]
\d B_{\m\n} & = & \ft12  \bar{\e} \g_{[\m} \p_{\n]} + C_{[\m} \d(\e)
C_{\n]}\ ,
\label{weylsusy}
\end{eqnarray}
where we have used the torsionful spin connection
\bea
{\widehat\omega}_\mu{}^{ab}_\pm &=& {\widehat\omega}_\mu{}^{ab} \pm {\widehat
H}_\mu{}^{ab}\ ,
\label{torsion}\w2
\widehat{\omega}^{ab}_\mu
&=& 2 e^{\nu[a} \partial_{[\mu} e_{\nu]}^{~b]} - e^{\nu[a} e^{b]\s} e_{\mu c}
\partial_\nu e^{~c}_\s +\ft12 \bar{\psi}^{[a} \gamma^{b]} \psi_\mu + \ft14
\bar{\psi}^a \gamma_\mu \psi^b \ \label{ohat} \eea
and the supercovariant curvatures
\bea
{\widehat\psi}_{\mu\nu} &=& 2D_{[\mu} ({\widehat\omega}_-) \psi_{\nu]} +i\g^\lambda
\widehat G_{\lambda[\mu}\psi_{\nu]}\ ,
\label{psiab}\w2
\widehat{G}_{\m\n} &=& 2\partial _{[\mu }C_{\nu]} + \ft12\rmi \bar{\p}_{[\m} \p_{\n]}\ ,
\label{ghat}\w2
\widehat{H}_{\m\n\r} &=& 3\partial _{[\mu }B_{\nu \rho ]} - \ft34
\bar{\p}_{[\m} \g_\n \p_{\r]} + \ft32 C_{[\m} G_{\n\r]}\ .
\label{hhat}
\eea
For the purposes of the next section we also define the supercovariant curvature
\bea {\widehat V}_{\mu\nu}{}^{ij} &=&
2\partial_{[\mu}V_{\nu]}{}^{ij} -2V_{[\mu}{}^{k(i}V_{\nu]k}{}^{j)} -
{\bar\psi}_{[\mu}^{(i}\gamma^\rho {\widehat\psi}_{\nu]\rho}^{j)}
\nn\w2 && +\ft1{6} {\bar\psi}_\mu^{(i} \gamma\cdot {\widehat
H}\psi_\nu^{j)} +\ft14 i {\bar\psi}_\mu^{(i} \gamma\cdot {\widehat
G} \psi_\nu^{j)}\ . \label{vmn} \eea

The redefinitions \eqref{redefs} are in fact equivalent to fixing the dilatation,
special conformal transformations and  $S$-supersymmetry  by imposing the gauge conditions
\begin{equation}\label{gc}
    \sigma=1\ , \qq b_\mu=0\ , \qq  \psi^i=0\ .
\end{equation}
The first of these conditions fixes the dilatation symmetry, the
second fixes the special conformal transformations, while the last
condition fixes the $S$-supersymmetries. We stress that the gauge fixing performed here is merely a way to describe a field redefinition and will not be used to obtain an off-shell Poincar\'e supergravity theory.
In fact, for that purpose a new compensating multiplet, which has been taken to be a linear multiplet in \cite{Bergshoeff:2002qk}, is needed.

In order for the condition $\psi^i = 0$ to be invariant under supersymmetry, one has to modify
the supersymmetry rules by adding a compensating $S$-supersymmetry
transformations with parameter
\begin{equation}\label{eta}
\eta^i= \left( -\g\cdot T +\frac14 \g\cdot {\widehat G}\right) \e^i \ .
\end{equation}
The second gauge condition in (\ref{gc}), in turn, leads to the
compensating conformal boost transformations with parameter
\begin{equation}\label{K}
\Lambda_{K\mu} = -\frac14 \rmi {\bar\e} \phi_\mu -\frac14 \rmi {\bar\eta}\psi_\mu
+\bar\e \g_\mu \chi \ ,
\end{equation}
with $\eta$ as given in (\ref{eta}).

Similarly, we change basis for the Yang-Mills multiplet by defining the dilatation and $S$-supersymmetry invariant tilded fields as
\begin{eqnarray}
{\tilde A}_\mu^I &=& A_\mu^I\ , \nn\w2 {\tilde Y}^{ijI}
&=&\sigma^{-2} Y^{ijI} +
\tfrac{1}{10}\sigma^{-2}{\bar\psi}_\mu^{(i}\gamma^\mu\lambda^{j)I}
-\ft1{20}i\sigma^{-2}\rho^I{\bar\psi}_\mu^{(i}\psi^{j)\mu}\ , \nn\w2
{\tilde \lambda}^{iI} &=&\sigma^{-3/2}\lambda^{iI}
-\ft1{5}i\sigma^{-3/2}\rho^I\gamma^\mu\psi_\mu^i\ , \nn\w2
{\tilde\rho}^I &=& \sigma^{-1}\rho^I\ .
\end{eqnarray}
Again, dropping the tildes for convenience in notation, we find the supersymmetry transformation rules
\begin{eqnarray}
\d A_\m^I &=& -\ft12\rmi \rho^I \bar{\e} \p_\m + \ft12 \bar{\e}
\g_\m \lambda^I \ ,
\nonumber \w2
\d Y^{ij\, I} &=& -\ft12 \bar{\e}^{(i} \widehat{\slashed{D}} \lambda^{j) I} -
\tfrac{1}{24} \bar{\e}^{(i} \g \cdot \widehat{H} \lambda^{j)I} -
\ft12 i g \bar{\e}^{(i} f_{JK}{}^I \rho^J \lambda^{j)K} \ ,
\nonumber \w2
\d \lambda^{i I} &=& - \ft14 \left(\g \cdot \widehat{F}^I-\rho^I \g\cdot \widehat{G}\right)
\e^i -\ft12\rmi \widehat{\slashed{D}}\rho^I \e^i - Y^{ij\, I} \e_j \ ,
\nonumber \w2 \d \rho^I &=& \ft12
\rmi \bar{\e} \lambda^I \ , \label{vectPoincare}
\end{eqnarray}
where $\widehat{F}_{\m\n}^I$ and ${\widehat D}_\mu\, \rho^I$ are as defined in \eqref{hatF} and \eqref{deriv1}, respectively, and
\bea
{\widehat D}_\mu \lambda^{i I} &=&
(\partial_\mu + \ft14\,  \widehat{\o}_\mu
{}^{ab}\g_{ab} ) \l^{iI} - V_\mu^{ij} \l_j^I + g f_{JK}{}^I A_\mu^J
\lambda^{iK}
\nn\\
&& + \ft14 \left( \g
\cdot {\widehat{F}}^I - \rho^I \g\cdot {\widehat G}\right) \p_\mu^i
+ \ft12\rmi \widehat{\slashed{D}}\rho^I  \p_\mu^i  + Y^{ij\, I} \p_{\mu\, j}\ .
\eea
With these results at hand, we are ready to make a connection
between the Dilaton Weyl multiplet and Yang-Mills multiplet
transformation rules.


\section{The  Weyl multiplet as a Yang-Mills Multiplet}


In this section, we will show that the following multiplet of fields

\begin{equation}
\left(\widehat{\o}_{\m}{}^{ab}_+,\
-\widehat{\psi}^i_{ab}, \ -{\widehat V}_{ab}{}^{ij},\  \widehat{G}_{ab}\right),
\end{equation}
defined in \eqref{torsion}, \eqref{psiab}, \eqref{vmn} and \eqref{ghat}, respectively,
transforms as a Yang-Mills multiplet

\begin{equation}
\left(A^I_\m,\ \l^{i I},\ Y^{ij\, I}, \ \rho^I \right),
\end{equation}
where the antisymmetric index pair $ab$ plays the role of the
Yang-Mills index $I$, for Yang-Mills group $\text{SO}(4,1)$.  In the
above the definition of  the gravitino curvature that follows from
\eqref{psiab} is given by
\be
{\widehat\psi}_{ab} = 2 D_{[a} (\widehat\omega, {\widehat\omega}_-) \psi_{b]}
+i \gamma^c {\widehat G}_{c[a}\psi_{b]}\ ,
\label{crucial}
\ee
where it is important to note that in $D_a (\widehat\omega, {\widehat\omega}_-) \psi_b$,
the connection ${\widehat\o}$ rotates the Lorentz vector index, while the
connection ${\widehat\omega}_-$ rotates the Lorentz spinor index.

Next, we calculate the transformation rules of
$\widehat{\o}_{\m}{}^{ab}_+$ and $\widehat{G}_{ab}$. In the new basis, we find
\begin{eqnarray}
\delta \widehat{\omega}_\mu{}^{ab} &=& -\ft12 \bar\e
\g^{[a}\widehat{\psi}_\mu{}^{b]}
-\ft14\bar\e\g_\mu\widehat{\psi}^{ab} -\ft12 \bar\e\g^c \psi_\mu
\widehat H_{cab}-\ft12\rmi\bar\e\psi_\mu \widehat G_{ab}\ , \nn\w2
\delta\widehat H_{\mu ab} &=&  -\frac12 \bar\e
\g_{[a}\widehat{\psi}_{b]\mu}
-\frac14\bar\e\g_\mu\widehat{\psi}_{ab} +\frac12 \bar\e\g^c \psi_\mu
\widehat H_{cab}\ , \w2 \delta \widehat G_{ab} &=&  -\frac12 \rmi
\bar\e \widehat{\psi}_{ab}\ .
\end{eqnarray}
From the first two equations it readily follows that
\be \delta \widehat{\omega}_\mu{}^{ab}_+ = -\frac12\rmi\widehat
G^{ab}\bar\e\psi_\mu -\ft12\bar\e\g_\mu\widehat{\psi}^{ab}\ .
 \ee
Comparing these results with the transformation rules of the
Yang-Mills multiplet, one sees that they indeed agree upon making
the identification
\begin{equation} \label{ident1}
\widehat{\o}_{\m}{}^{ab}_+ \ \ \leftrightarrow \ \ A_\m^I \ ,
\qquad {\widehat\psi}^{i\,ab} \ \ \leftrightarrow \ \ -\l^{i I}\ , \qquad \widehat{G}^{ab} \ \
\leftrightarrow \ \ \rho^I \ .
\end{equation}
Next, we compute the transformation rule for ${\widehat\psi}_{ab}^i$. We find
\begin{equation}
\d {\widehat\psi}^i_{ab} = \tfrac14 \widehat{R}_{abcd} ({\widehat
\omega}_-) \g^{cd} \e^i - {\widehat V}_{ab}{}^{ij}\e_j - \rmi
\widehat{D}_{[a}({\widehat\o},\widehat{\o}_-) \widehat{G}_{b]c} \g^c
\e^i + \tfrac12 \widehat{G}_{ca} \widehat{G}_{bd} \g^{cd} \e^i\ ,
\end{equation}
where ${\widehat R}_{abcd} ({\widehat\omega}_-)$ denotes the
super-covariant curvature of the torsionful connection
${\widehat\o}_-$, and in $\widehat{D}_a({\widehat\o},\widehat{\o}_-)
\widehat{G}_{bc}$  it is important to note that the connection
${\widehat\o}$ rotates the Lorentz vector index $b$, while the
connection ${\widehat\o}_-$ rotates the index $c$. This follows from
\eqref{crucial}. Next, using the Bianchi identity for $\widehat{H}$
\begin{equation}
\widehat{D}_{[a} (\widehat\o)\widehat{H}_{bcd]} = \ft34
\widehat{G}_{[ab} \widehat{G}_{cd]} \ ,
\end{equation}
one finds that
\begin{equation}
{\widehat R}_{abcd}({\widehat\o}_-) = {\widehat
R}_{cdab}({\widehat\o}_+) - \left(\widehat{G}_{ab} \widehat{G}_{cd}
+ 2 \widehat{G}_{a[c} \widehat{G}_{d]b} \right)\,.
\end{equation}
If one furthermore uses the Bianchi identity ${\widehat
D}_{[a}(\widehat{\o}) \widehat{G}_{cd]} = 0$, one finds the final
result
\begin{equation}
\d {\widehat\psi}^i_{ab} = \tfrac14 \g^{cd} {\widehat R}_{cdab}
({\widehat\omega}_+) \e^i - {\widehat V}_{ab}{}^{ij}\e_j + \ft12 i
\gamma^\mu \widehat{D}_\mu (\widehat{\o}_+) \widehat{G}_{ab} \e^i -
\tfrac14 \widehat{G}_{ab} \g \cdot \widehat{G} \e^i \ ,
\end{equation}
where in $\widehat{D}_\mu (\widehat{\o}_+)
\widehat{G}_{ab}$, the connection ${\widehat\o}_+$ rotates both of the indices $a$ and $b$. Upon using the identifications \eqref{ident1}, one sees that this transformation rule indeed assumes the form of
$\d \l^{i I}$, see \eqref{vectPoincare}, if one makes the extra identification
\begin{equation} \label{ident2}
{\widehat V}_{ab}{}^{ij} \ \ \leftrightarrow \ \ - Y^{ij\, I} \,.
\end{equation}

Finally, we calculate the transformation rule of ${\widehat V}_{ab}{}^{ij}$ in a similar way. We find
\be \d {\widehat V}_{ab}{}^{ij}  =  {\bar\epsilon}^{(i} \gamma^c
\widehat{D}_{[a}({\widehat\o},{\widehat\omega}_-)\psi_{b]c} -
\tfrac16 \bar{\e}^{(i} \g \cdot \widehat{H} {\widehat\psi}^{j)}_{ab}
- \ft14 i \bar{\e}^{(i} \g \cdot \widehat{G}
{\widehat\psi}^{j)}_{ab}\ , \ee
where in $\widehat{D}_a({\widehat\o},{\widehat\omega}_-)\psi_{bc}$
the connection ${\widehat\o}$ acts on the index $b$ while the
connection ${\widehat\o}_-$ acts on the index $c$ and the spinor
index. Upon using the Bianchi identity
\begin{equation}
\widehat{D}_{[a}(\widehat\o) \widehat{\psi}^i_{bc]} = -\ft16 i \g^d
\left(2 \widehat{G}_{d[a} {\widehat\psi}^i_{b]c} + \widehat{G}_{dc}
{\widehat\psi}^i_{ab} \right)\ ,
\end{equation}
we then find
\begin{equation}
\d {\widehat V}_{ab}{}^{ij} = -\tfrac12 \bar{\e}^{(i}
\slashed{\widehat{D}}({\widehat\o},{\widehat\o}_-)
{\widehat\psi}^{j)}_{ab} - \tfrac16 \bar{\e}^{(i} \g \cdot
\widehat{H} {\widehat\psi}^{j)}_{ab} - \rmi \bar{\e}^{(i}
\widehat{G}^d{}_{[a} {\widehat\psi}_{b]d}^{j)} \ , \label{deltaRV}
\end{equation}
where, in
$\widehat{D}_\mu({\widehat\o},{\widehat\o}_-){\widehat\psi}^{j)}_{ab}$
the connection ${\widehat\o}$ rotates the spinor index, while the
connection ${\widehat\o}_-$ rotates the Lorentz vector indices. The
expression \eqref{deltaRV} can equivalently be written as
\begin{equation}
\d {\widehat V}_{ab}{}^{ij}(V) = -\tfrac12 \bar{\e}^{(i}
\slashed{\widehat{D}}({\widehat\o}, {\widehat\o}_+)
{\widehat\psi}^{j)}_{ab} - \tfrac{1}{24} \bar{\e}^{(i} \g \cdot
\widehat{H} \psi^{j)}_{ab} - \rmi \bar{\e}^{(i} \widehat{G}^d{}_{[a}
{\widehat\psi}_{b]d}^{j)} \ ,
\end{equation}
where in $\widehat{D}_c({\widehat\o}, {\widehat\o}_+)
{\widehat\psi}_{ab}$ the connection ${\widehat\o}$ acts on the
spinor index, while ${\widehat\omega}_+$ acts on both of the indices
$a$ and $b$. This result indeed agrees with the corresponding
Yang-Mills transformation rule, upon using the identifications
\eqref{ident1} and \eqref{ident2}.

Summarizing, we find the correspondence
\vskip .2truecm
\begin{equation}\left(
                  \begin{array}{c}
                    A_\mu^I \\
                    \\
                    Y^{ij}_I \\
                    \\
                    \lambda^i_I \\
                    \\
                   \rho_I \\
                  \end{array}
                \right)    \qq \longleftrightarrow \qq
                 \left(
                  \begin{array}{c}
                    {\widehat\omega}_\mu{}^{ab}_+ \\
                    \\
                   - {\widehat V}_{ab}{}^{ij} \\
                    \\
                   - {\widehat\psi}_{ab}^i \\
                    \\
                   \widehat G_{ab} \\
                  \end{array}
                \right) \,.
\label{rules}
\end{equation}
\vskip .2truecm

This concludes our discussion of the analogy between the Dilaton
Weyl multiplet and the Yang-Mills multiplet. After constructing the
coupling of the Yang-Mills multiplet to the Weyl multiplet by means
of dimensional reduction from $6D$ in the next section, we shall use
the correspondence \eqref{rules} to obtain the Riemann squared
invariant in the subsequent section.


\section{ Yang-Mills Coupled to Weyl in $6D$ and Dimensional Reduction to $5D$}


In this section we will verify that the $5D$ local supersymmetry transformations of the
Weyl multiplet and Yang-Mills multiplet given above follow precisely from a suitable circle reduction and truncation of the known counterparts in $6D$ \cite{Bergshoeff:1985mz}. Next, we will reduce the action in $6D$ that describes the coupling of a Yang-Mills multiplet to the Weyl multiplet \cite{Bergshoeff:1985mz} down to $5D$.  Using these results, we shall then construct in section 6 the new supersymmetric Riemann tensor squared invariant by making use of  the analogy between the nonabelian vector multiplet and the Dilaton Weyl multiplet, derived in the previous section.


\subsection{The $6D$ Action for Yang-Mills Coupled to Weyl}


The $6D$ Weyl multiplet with the superconformal symmetries
gauge-fixed, or equivalently with suitable field redefinitions
amounting to the same thing, consists of the vielbein $E_M{}^A$,
gravitino $\Psi_M^i$, the $SU(2)$ valued vector fields ${\cal
V}_M^{ij}$ and the $2$-form potential ${\cal B}_{MN}$. Their
supersymmetry transformations are given by \cite{Bergshoeff:1986wc}
\bea
E_B{}^M\delta E_M{}^A &=& \ft12 {\bar\varepsilon}\Gamma^A\Psi_B\ ,
\nn\w2
\delta \Psi_A &=& {\cal D}_A (\widehat\Omega) \varepsilon + \ft18 \widehat{\cal H}_A{}^{BC}\Gamma_{BC}\varepsilon
-\left(E_A{}^M\delta E_M{}^B\right) \Psi_B\ ,
\nn\w2
\delta {\cal V}_A^{ij} &=& {\bar\varepsilon}^{(i} \Gamma^B \Psi_{AB}^{j)} -\ft16 {\bar\varepsilon}^{(i} \Gamma^{BCD}\Psi_A^{j)} \widehat{\cal H}_{BCD} -\left(E_A{}^M\delta E_M{}^B\right) {\cal V}_B^{ij}\ ,
\nn\w2
\delta {\cal B}_{AB} &=& -{\bar\varepsilon} \Gamma_{[A} \Psi_{B]}
+ 2\left(E_{[A|}{}^M\delta E_M{}^C\right) {\cal B}_{|B]C}\ ,
\eea
where
\bea
{\widehat\Psi}_{AB}^i &=& 2D_{[A} (\widehat\Omega) \Psi_{B]}^i +\ft14 \widehat{\cal H}_{CD[A} \Gamma^{CD}\Psi_{B]}^i +T_{AB}{}^C \Psi_C^i\ ,
\nn\w2
\widehat{\cal H}_{ABC} &=& 3\partial_{[A} {\cal B}_{BC]} +\ft32 {\bar\Psi}_{[A}\Gamma_B\Psi_{C]}
-3 T_{[AB}{}^D {\cal B}_{C]D}\ ,
\nn\w2
T_{AB}{}^C &=& E_A{}^M E_B{}^N \left(\partial_M E_N{}^C - \partial_N E_M{}^C\right)\ ,
\label{defs1}
\eea
and
\bea
{\widehat\Omega}_C{}^{AB} &=& 2E_C^M E^{N[A} \partial_{[M} E_{N]}{}^{B]} -E^{MA} E^{NB} \partial_{[M} E_{N]C}
+ \ft12 {\bar\Psi}_C \Gamma^{[A}\Psi^{B]} +\ft14 {\bar\Psi}^A\Gamma_C\Psi^B\ ,
\nn\w2
D_A (\widehat\Omega)\varepsilon_i &=& \partial_A \varepsilon^i + \ft14 {\widehat\Omega}_A{}^{BC} \Gamma_{BC}\varepsilon^i
+\ft12{\cal V}_A^{ij}\varepsilon_j\ .
\eea
Turning to the $6D$ Yang-Mills multiplet, it consists of a vector
$W_M$, a spinor $\Omega^i$ and a triplet of auxiliary fields ${\cal
Y}_{ij}$. The superconformal gauge fixed Yang-Mills supermultiplet
transformations take the form \cite{Bergshoeff:1986wc}
\bea
\delta W_A &=& -{\bar\varepsilon} \Gamma_A \Omega -\left(E_A{}^M\delta E_M{}^B\right) W_B\ ,
\nn\w2
\delta\Omega^i &=& \ft18 \Gamma^{AB} \widehat {\cal F}_{AB}\varepsilon^i - \ft12 {\cal Y}^{ij} \varepsilon_j\ ,
\nn\w2
\delta {\cal Y}^{ij} &=& -{\bar\varepsilon}^{(i}\Gamma^A \widehat{\cal D}_A \Omega^{j)} -\ft1{24} {\bar\varepsilon}^{(i} \Gamma^{ABC}\Omega^{j)} \widehat{\cal H}_{ABC}\ ,
\eea
where $W_A = W_A^I T_I$, and similarly for the other members of the Yang-Mills multiplet, where $T_I$ are the generators of the Yang-Mills gauge group, and
\bea
\widehat {\cal F}_{AB} &=& 2 \partial_{[A} W_{B]} +g [W_A,W_B] + 2{\bar\Psi}_{[A} \Gamma_{B]}\Omega
+T_{AB}{}^C W_C\ ,
\w2
\widehat{\cal D}_A\Omega^i &=& \partial_A \Omega^i + \ft14 {\widehat\Omega}_A{}^{BC} \Gamma_{BC}\Omega^i
+\ft12{\cal V}_A^{ij}\Omega_j -g[W_A,\Omega^i]
\nn\w2
&& -\ft18 \Gamma^{BC} \widehat{\cal F}_{BC}\Psi_A^i +\ft12 {\cal Y}^{ij}\Psi_{Aj}\ .
\eea

Finally, the locally supersymmetric Lagrangian in $6D$ that describes the couplings of the superconformal gauge fixed Yang-Mills multiplet and Weyl multiplet is given by \cite{Bergshoeff:1985mz}
\bea
E^{-1} {\cal L}_{6} &=& -\ft14 {\cal F}_{AB}^I {\cal F}_I^{AB} -2 {\bar\Omega} \Gamma^A D_A \Omega +{\cal Y}_{ij}^I{\cal Y}^{ij}_I
-\ft1{16} \varepsilon^{ABCDEF} {\cal B}_{AB}{\cal F}_{CD}^I {\cal F}_{EF}^I
\nn\w2
&& +\ft14 \left( {\cal F}_{AB}^I + \widehat{\cal F}_{AB}^I\right) {\bar\Omega}^I\Gamma^C\Gamma^{AB} \Psi_C
+\ft1{12} \widehat{\cal H}_{ABC} {\bar\Omega}^I\Gamma^{ABC}\Omega^I\ ,
\label{6DLag}
\eea
where $F_{\mu\nu}^I$ is the ordinary Yang-Mills field strength and
\be
{\cal D}_A \Omega^i = \partial_A \Omega^i + \ft14 {\widehat\Omega}_A{}^{BC} \Gamma_{BC}\Omega^i
+\ft12{\cal V}_A^{ij}\Omega_j -g[W_A,\Omega^i]\ .
\ee


\subsection{Dimensional Reduction to $5D$ }


We begin by making the ansatz\footnote{ Due the sign in the relation
between ${\cal B}_{ab}$ and $B_{ab}$, the torsionful spin connection
${\widehat\o}_-$ in \cite{Bergshoeff:1986vy} corresponds to
${\widehat\o}_+$ here.}
\bea
E_M{}^A &=& \left(
           \begin{array}{cc}
             e_\mu{}^a & -C_\mu \\
             0 & 1 \\
           \end{array}
         \right)\ ,\qquad E_A{}^M =\left(
           \begin{array}{cc}
             e_a{}^\mu & e_a{}^\mu C_\mu \\
             0 & 1 \\
           \end{array}
         \right)\ ,
\nn\w4
{\cal B}_{ab} &=& -2 e_a{}^\mu e_b{}^\nu B_{\mu\nu}\ ,\qquad {\cal B}_{a5}= -e_a{}^\mu C_\mu\ ,\qquad {\cal V}_a^{ij}= -2 e_a{}^\mu V_\mu^{ij}\ ,\qquad {\cal V}_5^{ij}=0\ ,
\nn\w2
\Psi_a &=&e_a{}^\mu \psi_\mu\ ,\qquad \Psi_5=0\ ,\qquad \varepsilon = \epsilon\ ,
\nn\w4
W_a &=& e_a{}^\mu A_\mu\ ,\qquad W_5 = \rho\ , \qquad {\cal Y}^{ij}= -Y^{ij}\ ,
\qquad \Omega=-\ft12 \lambda\ .
\eea
We also let
\be
\Gamma_a = i\gamma_a\gamma_5\ ,\qquad \bar{\varepsilon} = i\bar{\epsilon} \gamma_5\ ,
\qquad \varepsilon^{abcde5} =\varepsilon^{abcde}\ .
\label{gammas}
\ee
The second expression in \eqref{gammas} applies to all Dirac conjugated spinors. Note also that we have identified the Kaluza-Klein vector originating from the $6D$ metric with ${\cal B}_{a5}$, and we have set to zero ${\cal V}_5^{ij}$ and $\Psi_5$. This amounts to a consistent truncation of a single off-shell vector multiplet in $5D$.

It is now a straightforward exercise to show that the above ansatz yields precisely the local supersymmetry transformations \eqref{fixed} and \eqref{vectPoincare}. In doing so, and in reducing the Lagrangian \eqref{6DLag} to $5D$, it is useful to note the relations
\bea
&& \widehat{\cal F}_{ab} = {\widehat F}_{ab}  -\rho {\widehat G}_{ab}\ ,
\qquad
\widehat{\cal F}_{a5} = {\widehat D}_a \rho\ ,
\nn\w2
&& \widehat{\cal H}_{abc} = -2{\widehat H}_{abc}\ ,\qquad\widehat {\cal H}_{ab5} = -{\widehat G}_{ab}\ ,
\nn\w2
&& {\widehat\Omega}_c{}^{ab}= {\widehat\omega}_c{}^{ab}\ ,
\qquad {\widehat\Omega}_5{}^{ab} =\ft12 {\widehat G}^{ab}\ ,\qquad
{\widehat\Omega}_a{}^{b5} = -\ft12 {\widehat G}_a{}^b\ ,
\eea
where $\widehat{F}_{\m\n}^I, \widehat{G}_{\m\n}, \widehat{H}_{\m\n\r}$ and $\widehat{\omega}^{ab}_\mu$
are defined in \eqref{hatF}, \eqref{ghat}, \eqref{hhat} and \eqref{ohat}, respectively.

It is also useful to record the results
\bea
D_a  \Omega &=& -\ft12 D_a(\widehat\omega) \lambda +\ft18 i {\widehat G}_{ab} \gamma^b \lambda\ ,
\nn\w2
D_5 \Omega &=& -\ft1{16} {\widehat G}_{ab} \gamma^{ab} \lambda\ .
\eea
Armed with these results, it is straightforward to reduce the Lagrangian \eqref{6DLag} to $5D$. The result is
as follows:
\bea e^{-1} {\cal L}_{5} &=& -\ft14 \left(F_{\mu\nu}^I-\rho^I
G_{\mu\nu}\right) \left( F^{\mu\nu I}-\rho^I G^{\mu\nu}\right)
-\ft12 D_\mu \rho^I D^\mu \rho^I -\ft12 {\bar\lambda}^I \slashed{D}
\lambda^I +Y_{ij}^I Y^{ij I} \nn\w2 &&
+\ft1{16}\varepsilon^{\mu\nu\rho\sigma\lambda} \left(F_{\mu\nu}^I
-\rho^I G_{\mu\nu}\right) \left[\, (F_{\rho\sigma}^I -\rho^I
G_{\rho\sigma}) C_\lambda+8 B_{\rho\sigma} D_\lambda \rho^I\right]
\nn\w2 && -\ft12 i {\bar\lambda}^I \gamma^\nu\gamma^\mu\psi_\nu
\left(D_\mu \rho^I-\ft14 i{\bar\psi}_\mu\lambda^I\right) \nn\w2 &&
-\ft14 \left(F_{\mu\nu}^I-\rho^I G_{\mu\nu} -\ft12 {\bar\psi}_\mu
\gamma_\nu\lambda^I\right) {\bar\lambda}^I
\gamma^\rho\gamma^{\mu\nu}\psi_\rho \nn\w2 && -\ft1{24} {\widehat
H}_{\mu\nu\rho} {\bar\lambda}^I\gamma^{\mu\nu\rho}\lambda^I -\ft1{8}
i {\widehat G}_{\mu\nu} {\bar\lambda}^I \gamma^{\mu\nu}\lambda^I\ .
\label{Lag5} \eea This Lagrangian will be our starting point for
constructing the supersymmetric Riemann tensor squared action in the
next section.

\section{The Riemann Squared Invariant}


To obtain the Riemann squared invariant, we  make the substitutions \eqref{rules} in the Lagrangian
\eqref{Lag5}. Thus we find the main result of this paper given by
\bea e^{-1}{\cal L}(R^2) &=& -\ft14 \left[\,R_{\mu\nu
ab}({\widehat\omega}_+)- G_{\mu\nu} {\widehat G}_{ab}\right]
\left[\,R^{\mu\nu ab}({\widehat\omega}_+)- G^{\mu\nu} {\widehat
G}^{ab}\right] \nn\w2 && -\ft12 D_\mu({\widehat\omega}_+) {\widehat
G}^{ab} D^\mu({\widehat\omega}_+){\widehat G}_{ab} +{\widehat
V}_{\mu\nu}{}^{ij} {\widehat V}^{\mu\nu}{}_{ij} -\ft12
{\widehat\psi}^{ab} {\slashed D} ({\widehat\o},{\widehat\omega}_+)
{\widehat\psi}_{ab} \nn\w2 && +\ft1{16}
\varepsilon^{\mu\nu\rho\sigma\lambda} \left(\,R_{\mu\nu
ab}({\widehat\omega}_+)- G_{\mu\nu} {\widehat G}_{ab}\right)
\left(\,R_{\rho\sigma}{}^{ab}({\widehat\omega}_+)- G_{\rho\sigma}
{\widehat G}^{ab}\right) C_\lambda \nn\w2 && +\ft12
\varepsilon^{\mu\nu\rho\sigma\lambda} \left(\,R_{\mu\nu
ab}({\widehat\omega}_+)- G_{\mu\nu} {\widehat G}_{ab}\right)
\left(D_\lambda ({\widehat\o}_+) G^{ab}\right)B_{\rho\sigma} \nn\w2
&& -\ft12 i {\bar\psi}_\nu \gamma^\mu\gamma^\nu {\widehat\psi}^{ab}
{\widehat D}_\mu({\widehat\o}_+) {\widehat G}_{ab} +\ft14
\left(\,R_{\mu\nu ab}({\widehat\omega}_+)- G_{\mu\nu} {\widehat
G}_{ab}\right) {\bar\psi}_\rho\gamma^{\mu\nu}\gamma^\rho
{\widehat\psi}_{ab} \nn\w2 && -\ft1{24} {\widehat H}_{\mu\nu\rho}
{\widehat{\bar\psi}}_{ab} \gamma^{\mu\nu\rho}{\widehat\psi}^{ab}
-\ft18 i {\widehat G}_{\mu\nu} {\widehat{\bar\psi}}_{ab}
\gamma^{\mu\nu}{\widehat\psi}^{ab} \nn\w2 && -\ft18 {\bar\psi}_\nu
\gamma^\mu\gamma^\nu {\widehat\psi}^{ab} {\bar\psi}_\mu
{\widehat\psi}_{ab} +\ft18 {\bar\psi}_\rho
\gamma^{\mu\nu}\gamma^\rho {\widehat\psi}^{ab} {\bar\psi}_\mu
\gamma_\nu{\widehat\psi}_{ab}\ . \eea
The action of this Lagrangian is invariant under the off-shell
${\cal N}=2, D=5$ supersymmetry transformations given in
\eqref{weylsusy}. The use of the Lorentz vector indices is motivated
by the substitution rule \eqref{rules}. As a consequence, in
$D_\mu({\widehat\o}_+) {\widehat G}_{ab}$ the spin connection
${\widehat \o}_+$ rotates the indices $a$ and $b$, while in $D_\mu
({\widehat\o},{\widehat\omega}_+) {\widehat \psi}_{ab}$ the
connection ${\widehat\o}$ rotates the spinor index, and the
connection ${\widehat\omega}_+$ rotates the indices $a$ and $b$.

The purely bosonic part of the Lagrangian takes the form
\bea \label{bosonic} e^{-1}{\cal L}(R^2)_{\rm bosonic} &=& -\ft14
\left[\,R_{\mu\nu ab}(\omega_+)- G_{\mu\nu} G_{ab}\right]
\left[\,R^{\mu\nu ab}(\omega_+)- G^{\mu\nu} G^{ab}\right] \nn\w2 &&
-\ft12 D_\mu(\omega_+) G^{ab} D^\mu(\omega_+)G_{ab}
+V_{\mu\nu}{}^{ij} V^{\mu\nu}{}_{ij} \nn\w2 && +\ft1{16}
\varepsilon^{\mu\nu\rho\sigma\lambda} \left(\,R_{\mu\nu
ab}(\omega_+)- G_{\mu\nu} G_{ab}\right)
\left(\,R_{\rho\sigma}{}^{ab}(\omega_+)- G_{\rho\sigma}
G^{ab}\right) C_\lambda \nn\w2 && +\ft12
\varepsilon^{\mu\nu\rho\sigma\lambda} \left(\,R_{\mu\nu
ab}(\omega_+)- G_{\mu\nu}G_{ab}\right) \left(D_\lambda (\omega_+)
G^{ab}\right)B_{\rho\sigma}\ . \eea

It is possible to extend the above  result by adding the
Hilbert-Einstein term, as well as the Weyl squared invariant of
\cite{Hanaki:2006pj}. To do so, one first performs the inverse of
the field redefinitions \eqref{redefs} making the fields $\sigma$
and $\psi^i$ explicit.  Next, the Weyl squared invariant of
\cite{Hanaki:2006pj}, prior to any conformal symmetry gauge fixing,
can be added to our action. The Standard Weyl multiplet used in that
action can be converted to the Dilaton Weyl multiplet by using the
map that exists between these two multiplets, see eqs.
\eqref{TchiindilW} and \eqref{valueD}. Finally, a superconformal
version of the Einstein action, using the linear multiplet as a
compensating multiplet, can be added to these two actions. A
conformal gauge-fixing at the very end then leads to the desired
result. Alternatively, instead of giving the Einstein actoin a
superconformal treatment, one can also reduce the off-shell
Poincar\'e supergravity constructed in
\cite{Bergshoeff:1985mz,Coomans:2011ih} to $5D$ in a manner
described in this work.

We note that in the off-shell Poincar\'e supergravity theory the
vector fields $V_\mu{}^{ij}$ are auxiliary. However, with the
addition of our Riemann squared invariant, these fields acquire
kinetic terms and become dynamical. Such dynamical auxiliary fields
should be treated with care in a string theory
approximation.\,\footnote{We thank Bernard de Wit for a clarifying
discussion on this point.}


\section{Conclusions}


In this note we have constructed a new $D=5, \mathcal{N}=2$
supersymmetric Riemann tensor squared action. A noteworthy feature
of this action is that it contains the purely gravitational
Chern-Simons action \eqref{pure}. This is in contrast to the higher
order invariant constructed in \cite{Hanaki:2006pj} whose leading
term is the Weyl tensor squared and which contains the mixed
Chern-Simons term \eqref{mixed}. The latter invariant plays an
important  role in higher-order considerations in black hole entropy
calculations and the AdS/CFT correspondence. We expect our newly
constructed invariant to play a similar role too. In particular, it
would be interesting to see whether our new action may lead to
higher-order corrections to the entropy formula of black holes in a
similar way as the higher-order invariant of \cite{Hanaki:2006pj}
did \cite{review}. In this context, it is of interest to note that a
similar torsionful ${\rm Riem}^2$-invariant in three dimensions did
not lead to any correction of the central charge of the
corresponding boundary conformal field theory due to a curvature
with paralellizing torsion \cite{Andringa:2009yc}. For general
$R^2$-invariants, however, one does expect corrections, see
e.g.~\cite{Cvitan:2007en}.

Our construction was based on the methods developed some time ago in
the context of $D=6$ dimensions
\cite{Bergshoeff:1986vy,Bergshoeff:1986wc,Bergshoeff:1987rb}. Both
in $D=6$ and $D=5$ dimensions use is made of the observation that
the underlying Weyl multiplet contains a dilaton scalar field which
acts as the compensating field for dilatations. For this reason this
multiplet was nominated the Dilaton Weyl multiplet. It turns out in
a special basis where all fields are inert under dilatations and
S-supersymmetry  the Dilaton Weyl multiplet transforms precisely as
a Yang-Mills multiplet whose action can easily be constructed. This
is what makes the construction of the $D=5$ Riemann tensor squared
invariant feasible.

The maximally symmetric vacuum solutions and the resulting spectrum
corresponding to the ${\rm Riem}^2$ action \eqref{bosonic}  remain
to be investigated. The field redefinitions \eqref{redefs} make the
model invariant under the superconformal transformations
\eqref{weyl}, similar to a Brans-Dicke type realization of a
conformal Einstein action. Therefore, the conformal symmetries  may
be viewed as a ``fake'' symmetry in a sense. Consequently, whether
the formulation of the theory in this set up can lead to the
possibility of discarding potentially ghostly states in a manner
proposed in \cite{Maldacena:2011mk}, and investigated further in
\cite{Lu:2011ks,Lu:2011mw}, remains to be studied. In this context,
it is of interest to note that in any dimension $D$  the Riemann
tensor squared (${\rm Riem}^2$) can be written as the sum of a Weyl
tensor squared ($C^2$) and a fourth-order derivative action for a
compensating scalar $\phi$, both of which have been considered in
\cite{Maldacena:2011mk}, as follows \cite{Bergshoeff:1986wc}

\begin{eqnarray}
R_{\mu\nu}{}^{ab}R_{\mu\nu}{}^{ab} &=&
\phi^{D-4}C_{\mu\nu}{}^{ab}C_{\mu\nu}{}^{ab} + (D-2)\phi^{D-2}({\cal
D}_\mu\partial_\nu\phi^{-1})({\cal
D}_\mu\partial_\nu\phi^{-1})\nonumber\\[.2truecm]
&&+ 4\phi^{D-2}({\cal D}^\lambda\partial_\lambda\phi^{-1})^2
-8(D-1)\phi^{D-1}({\cal
D}^\lambda\partial_\lambda\phi^{-1})(\partial_\mu\phi^{-1})^2\\[.2truecm]
&&+
2D(D-1)(\partial_\nu\phi^{-1})^2(\partial_\nu\phi^{-1})^2\,.\nonumber
\end{eqnarray}

Finally, following \cite{Lu:2010ct} the compactification of the new
$R+R^2$ invariant  over $S^2$ to $D=3$ dimensions  is expected to
yield, after truncation, a supersymmetric version of topological
massive gravity. It would be interesting to explicitly perform this
reduction.

\subsection*{Acknowledgments}

E.S. would like to thank Groningen University for hospitality where
part of this work was done. The research of E.S. is supported in
part by NSF grants PHY-0555575 and PHY-0906222. The work of J.R. is
supported by the Stichting FOM.

\newpage

\providecommand{\href}[2]{#2}\begingroup\raggedright

\endgroup

\end{document}